# Knowledge Discovery in the SCADA Databases Used for the Municipal Power Supply System

Valery Kamaev[1], Alexey Finogeev[2], Anton Finogeev[3], Sergey Shevchenko[4]

[1]Volgograd State Technical University, Volgograd, Russia,
`kamaev@cad.vstu.ru`,
[2]Penza State University, Penza, Russia,
`alexeyfinogeev@gmail.com, fanton3@yandex.ru,`
[3]National Technical University "Kharkiv Polytechnic Institute", Kharkiv, Ukraina,
`sv_shevchenko@mail.ru`

**Abstract.** This scientific paper delves into the problems related to the development of intellectual data analysis system that could support decision making to manage municipal power supply services. The management problems of municipal power supply system have been specified taking into consideration modern tendencies shown by new technologies that allow for an increase in the energy efficiency. The analysis findings of the system problems related to the integrated computer-aided control of the power supply for the city have been given. The consideration was given to the hierarchy-level management decomposition model. The objective task targeted at an increase in the energy efficiency to minimize expenditures and energy losses during the generation and transportation of energy carriers to the Consumer, the optimization of power consumption at the prescribed level of the reliability of pipelines and networks and the satisfaction of Consumers has been defined. To optimize the support of the decision making a new approach to the monitoring of engineering systems and technological processes related to the energy consumption and transportation using the technologies of geospatial analysis and Knowledge Discovery in databases (KDD) has been proposed. The data acquisition for analytical problems is realized in the wireless heterogeneous medium, which includes soft-touch VPN segments of ZigBee technology realizing the 6LoWPAN standard over the IEEE 802.15.4 standard and also the segments of the networks of cellular communications. JBoss Application Server is used as a server-based platform for the operation of the tools used for the retrieval of data collected from sensor nodes, PLC and energy consumption record devices. The KDD tools are developed using Java Enterprise Edition platform and Spring and ORM Hibernate technologies. To structure very large data arrays we proposed to organize data in the form of hyper tables and to use the Compute Unified Device Architecture technology.

# 1 Introduction

An efficient and high-quality power supply for industrial companies is one of the main vectors of the development of the economy and successful advance of the contemporary society. The studies of the management processes of municipal power supply services that include electrical, thermal, water and gas supply engineering systems showed that the body of problems solved at different stages of decision-making can be considered in terms of system-synergetic approach [1].

The managerial structure used for the municipal power supply systems can be represented by the distribution of dedicated hierarchy –level control functions that form the clusters of problems according to the periodicity of their solution, functional content and the level of their importance. The managerial decision-making support is based on the realization of integrated methods used for the acquisition of many different data during the monitoring of the main facilities of municipal power supply system and technological processes. A theoretical statement of the problem related to an increase in the managerial efficiency takes into consideration the following features, in particular:

— realizing the strategy of optimal balancing for the solution of two problems: self-organization of efficient structures used for the off-line control and centralized coordination of the subsystems belonging to a lower level of the hierarchy;
— taking into consideration local priorities for the decision-making with regard to individual control subsystems;
— establishing optimal vertical links between the hierarchy levels of the subsystems of upper levels using the appropriate control actions;
— establish optimal horizontal links between subsystems one level using the exchange of structured information;
— optimization management functions of subsystems with different properties hierarchy of priority in accordance decision-making support.

# 2 Problems of the management automation by municipal power supply systems

Let's consider the problems faced by the developers of computer-aided decision-making support and control systems intended for the flow control services of the companies that maintain power supply networks [2].

1. Data "opacity" problem. It is known that most analytical information systems store their data and the results of work in the form, which is not appropriate for the use by other systems. The "opacity' of data protocols and formats does not allow the specialists of the company to transfer data to a different software environment or demands from specialists to apply every effort. Actually the company experiences "informational dependence" on the developer of a specific software product that may result in the disastrous consequences when the developer stops his engineering support.

2. Data mismatch problem. On the one part this problem is caused by the fact that power supply companies use different hardware, software and proprietary protocols for the data acquisition and processing. A proprietary character of the internal protocols of data exchange does not allow for the arrangement of data interrelation between the systems of different producers. For example, more than 60 types of heat computers with internal data exchange protocols are nowadays produced for municipal heat supply systems, which prevents the use of universal thermal energy control and record system for heat and hot water supply systems [3]. To solve this problem the technologies of integrated interface are used for the automated control of objects and technological processes [4] based on OPC (OLE for process control) standards. Though the introduction of appropriate standards into the SCADA systems for the acquisition of on-line (OPC data access) and archival (OPC Historical Data Access) data provided by industrial automatic systems is rather promising a lot of equipment and software still fails to use OPC technologies.

On the other part, the fact that the same data are required for the solution of the problems by different organizations that keep record of their own databases and use different software systems also presents a problem [5]. Moreover, the databases and programs present the same figures in different formats. For example, the section length of a heat network pipeline is used for hydraulic computations, thermal losses computations, accounting records of depreciation charges, repair scheduling, and the specification of the alienation zone in cadastral plans, etc. If the servicing company introduces changes into the section length due to the pipeline repair this will result in the disagreement of this figure in the databases of different organizations involved in the appropriate record-keeping. The synchronization of data in all databases requires considerable effort, labor time and appropriate expenses and very often this process is impossible to control. Therefore, it is impossible to provide reliable information data arrays, and this affects the efficiency and quality of engineering, technical and managerial decision making.

3. Information flow mismatch problem. Today engineering, technical, technological and organizational problems are solved using the transport telecommunication environment for the data exchange. Network technologies are called to solve the second problem of data disagreement in different organizations. However, lack of the uniform strategy for the use of network technologies, availability of multitude of telecommunication solutions, hardware and providers of network services result in the disagreement of information flows due to the implementation of different telecommunication solutions by companies.

4. Branch-specific data record problem [6]. This problem is related to the fact that municipal companies supplying energy of a different type (power supply, heat supply, water supply and gas supply) belong to different trade managerial proprietors that are mutually interrelated only at the level of mutual settlements of accounts. A comprehensive solution of this problem requires the introduction of the uniform SCADA system for all energy saving companies in the city, the arrangement of multidimensional "cloudy" storage of common data and the creation of the uniform transportation environment for the data exchange. A uniform monitoring and flow control system of engineering network facilities allows for the realization of the coordinated management of the industrial and technological processes of municipal power supply services. It also provides the consistency of the operation of all appropriate links, trouble-free operation of auxiliary, maintenance and trouble repair services.

These problems and lack of uniform strategy for their solution so much needed for the municipal power supply system does not allow people involved in decision making to see the entire patters of the processes that occur in the common municipal system of engineering communications, which reduces the managerial efficiency on the whole.

## 3   Arrangements made to provide the operation of intelligent analysis system

The dispatch systems SCADA are used to take measurements, collect data and analyze the parameters of energy carriers, process parameters used for their generation, transportation, consumption and utilization [7]. The purpose of the development of a comprehensive decision-making support and monitoring system for the municipal power supply system is to reach the power effectiveness through an increase in the power efficiency of power supply companies [8, reduction of power losses during the transportation of energy carriers, and the optimization of energy consumption in public and domestic buildings taking into consideration the meteorological information, information on building operation modes, the behavior of end users, etc.

The system used for the intelligent analysis and knowledge discovery in the stored data can operate under the condition of taking the following measures:

1.   Mounting energy consumption record devices with built-in OPC servers for engineering network facilities and end users;
2.   C Creating the wireless transport medium with sensor and cellular segments for the data acquisition [9];
3.   Providing the data collection using sensor units and cellular modems with automation devices via OPC servers;
4.   "Sinking" the data obtained from different sources into the multidimensional "cloudy" storage system supporting simultaneously consolidation, "refinement", normalization and transformation operations;
5.   Preparing analyzed data samples (including learning samples for forecast models), retrieving them from storages and other sources by different slices to transmit them to hypercubes for analytical processing;
6.   Synthesizing forecast models for the purpose of iterative selection of the best model to forecast energy consumption at different facilities depending on the factors available for short-term and long-term periods [10,11];
7.   Visualizing forecast results on digital cartographic map (DCM) that provides an opportunity for the display of the figures and performances of energy consumption, energy losses and energy efficiency in numerical and graphic forms using color-differentiation circuits;

8. Computing energy consumption parameters for different categories of consumers according to the forecast results

9. Transmitting the forecast results, energy consumption parameters in the form of structured reports, diagrams, and color circuits of DMP to dispatchers, energy managers and the managers of services and companies to work out measures that would allow for an increase in energy effectiveness;

10. Transfer the forecast results, the parameters of energy consumption in the form of structured reports, graphs, color schemes, DMP (dispatchers, energy managers, service managers and companies) to develop activities to improve energy efficiency;

11. Computer-aided control of the energy consumption and on-line correction of forecast parameters according to the real factors for the purpose of the iteration adjustment of forecast models and repeated forecasting [12].

## 4 Heterogeneous Wireless Medium for the Sensor Data Collection

Heterogeneous wireless medium for the collection of the sensor data to provide the operation of the system used for the intelligent analysis and knowledge discovery includes the following components:

1. Segments of ZigBee sensory networks with units connected to industrial automation devices, different sensors, fire alarm and burglar systems controlled by a segment coordinator;
2. Segments of Bluetooth networks [13] with units connected to industrial automation devices, different sensors, fire alarm and burglar systems controlled by master devices;
3. Cellular communication networks for the collection of data from remote objects if there is no opportunity to create own VPN circuits or it is not reasonable to do that from the economic standpoint;
4. Satellite network for the transport monitoring of company's vehicles and tracking mobile means used by the personnel for the communication;
5. WiFi and Ethernet segments of power supply facilities and dispatch centers;
6. Internet segment for remote access to the information resources of SCADA systems and monitoring systems.

To provide appropriate urban communication range it is recommended to use the following methods of data acquisition:

1. Placing and using the GSM/GPRS modems to retranslate data via the cellular network. The information can be transmitted by GSM-systems in the form of SMS-messages, via the modem connection (CSD), through the transmission of tone ringing (DTMF mode) and also in the GPRS mode of batch messages. The OPC server or PLC with intergrated modems are installed in the engineering network facilities and provide the acquisition, storage and processing of primary data provided by different measuring devices with the transmission of data to the dispatcher server through the

GSM/GPRS channels. However, the drawbacks of similar systems are low noise immunity, easy suppression of the GSM-channel, unstable operation of the GSM network, exposure of the network to different attacks, dependence on the operation of cellular network, and financial dependence on the service provider.

2. Placing and using the sensor-based units of the Zigbee network (Fig. 1) for engineering communication facilities with the possibility of the data acquisition and their transmission through the central network coordinator to the dispatcher server [14] using the following equipment::

  (a) External antennas with a high gain factor;
  (b) Low-power intermediate relays of the sensor network that can be installed:
    (i) on tall buildings and posts of power transmission lines;
    (ii) on external metering terminals of underground heat pipelines with the wire leakage control system that are arranged at a space of 300 meters from each other according to the specifications. In this case the repeater performs two functions: the frame relaying and leakage detection and localization.

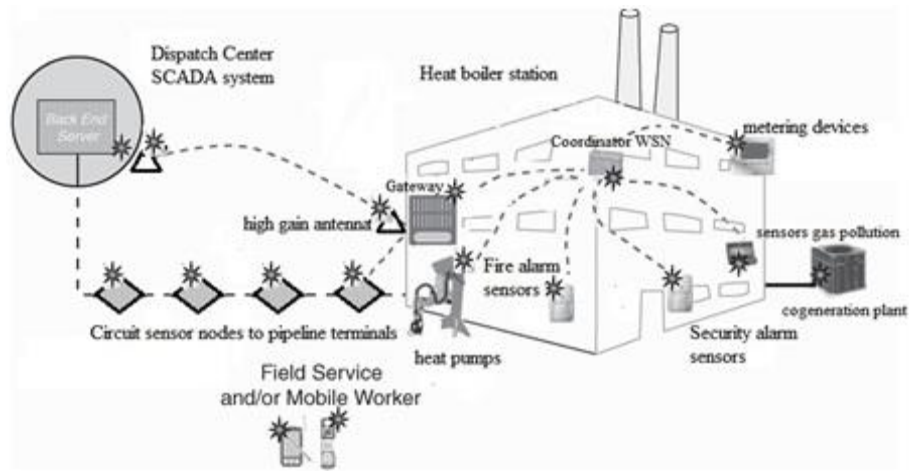

**Fig. 1.** - Sensor network cluster for heating points

The wireless sensor-based network can perfectly be used for SCADA systems to support the decision-making [15]. Such a network serves as an infrastructure for the collection of data provided by different measurement devices. This allows the dispatchers not only to perform monitoring and to read and analyze on-line and archive data, but also to transmit control actions to actuating mechanisms. The advantage of such a system consists in the possibility of the use of energy-saving modes when sensory units are mainly in the sleeping mode and turn on just to read the data and transmit them to dispatcher server. The use of a virtual corporate network for the data acquisition is also of great importance both from the financial point of view and from the point of view of providing the information security for the corporate computer-aided system that exercises control of technological processes for municipal power supply.

An interesting solution will be the arrangement of the relays of sensory network on external ground-based terminals that are connected to the leakage control system in contemporary heat mains. Contemporary heat-carrier transfer mains use double tubes with the internal heat insulation material and wired on-line remote control system (ORC). The connection to the wires of the ORC system is done using the instrument terminals that are connected to pipe conductors and are taken out to the surface. Intermediate terminals are arranges at a space of 300 meters according to specifications. The terminals can be used to provide data transfer from cluster-type sensory and other segments through the train of sensory units to dispatcher stations.

In the general case it is necessary to use all types of accessible wireless networks [16] to create the transport medium for the data acquisition and transmission from the monitoring facilities of the municipal power supply system that are distributed across the large city territory. This allows for an increase in the reliability due to the reservation of communication channels. Figure 2 shows the segment of such a heterogeneous wireless network for the data collection in the dispatch SCADA system used for the computer-aided control of municipal networks [17].

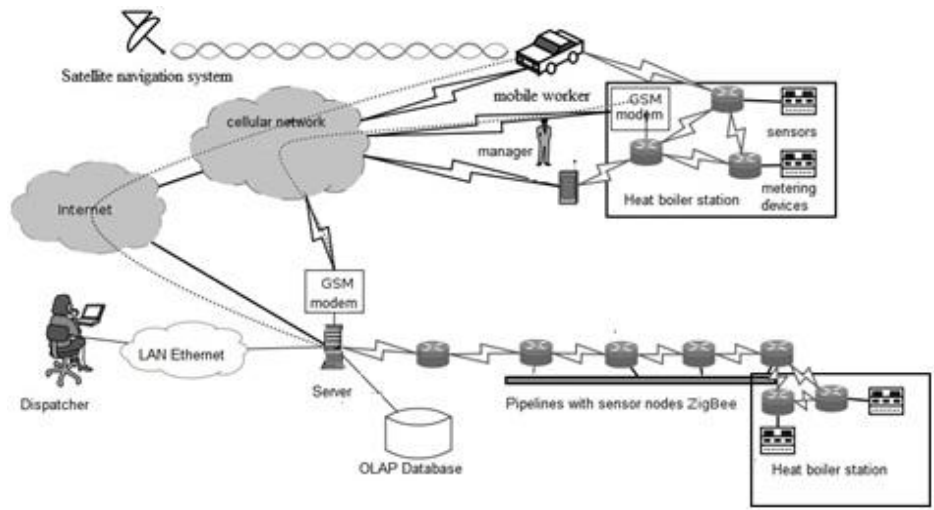

**Fig. 2.** – The wireless network segment for data collection of heating supply system

## 5   Data Analysis and Knowledge Discovery Tools in SCADA Databases Used for the Municipal Power Supply

SCADA systems receive on-line and archive data from OPC servers through the sensory network units or cellular modems and transmit them to the data preparation subsystem and their "immersion" into the multidimensional storage system.

As the number of monitoring objects and records in the multidimensional storage system increases the intelligent data analysis system will sooner or later encounter efficiency problems. These problems can be related to the improperly designed architecture system or can be caused by external restrictions. The latter can be presented by insufficient facilities of application servers or database servers, and nonoptimal structure of the databases. The first problem is solved through the addition of the servers to the cluster and the second problem is solved through the optimization of the data storage scheme, the transcription of the code of the intelligent analysis application, the segmentation of complicated relational inquiries, etc.

To increase the sampling efficiency of data slices from the multidimensional storage system we propose the data storage method, which is based on the combination of the industrial SQL storage system and distributed no relational data array storage system. This is the option for the solution of the problem related to the storage efficiency and scalability, because it is based on a simpler info logical data model. For this purpose the system of intelligent analysis in combination with the Oracle data manager uses the distributed no relational Cassandra system for caching the slices of multidimensional storage; this provides a significant increase in the data sampling rate, and improves its fault-tolerance and scalability.

The data scheme is described using the structures of hash tables, trees, etc. The Cassandra system using the Java platform includes the distributed hash system, providing thus the linear scalability with an increase in the data level. The data analysis and knowledge discovery system uses the data storage model in the form of the hyper table based on the family of columns, which differs from other similar systems that store data in a key/value pack. The hierarchy organization for the storage of caches with several nesting levels has also been realized. The scheme offered for the data storage and processing is attributed to the category of the storages that show an increased immunity to malfunctions because the data are self-replicated in the cluster "cloud" of the units of the distributed network.

The specificity of the operation of non relational database during the data acquisition from engineering power supply network facilities is that the data deletion and data change are not required. The data are only replenished as a rule in large blocks during the inquiry of OPC (OLE for Process Control) servers. Each individual record of the non relational component corresponds to the cached slice from the relational Oracle database. To optimize the efficiency the initial Cassandra code was changed in a special assembly, which uses data blocks of 32Mb reducing thus their number and increasing their sampling rate and retrieval rate.

The tools used for the intelligent data analysis and knowledge discovery operate on the side of the cloud of "servers" and are developed using Java Enterprise Edition (J2EE) platform complemented by the multilayer platform used for the development of corporate Spring Framework applications and the technology of the object-relational mapping (ORM) Hibernate.

The object-relational adapter (ORM) Hibernate is used to provide the flexibility of inquiries and the storage –related operation transparency. Particularly Hibernate uses the Cassandra system as the intermediate layer (level-two cache) between the intelligent analysis application and the relational database. Thanks to such an approach we managed to combine the advantages of relational and non relational data storage systems and increase the data analysis efficiency. In this system the Hibernate library used for the solution of object-relational design problems solves the problem of associations of Java classes with database tables and Java data types with SQL data types and provides also tools for automatic generation and updating hype table columns, and also for the arrangement of inquiries and processing of the obtained data.

The data analysis and knowledge discovery system used the JBoss Application server with the public source code as a server platform. HTTPS и AMF (Adobe Media Format) protocols were selected to exchange data between the client applications and the servers and call to remote server procedures of the business-logic. The user's interface is realized using the Adobe Flex platform, which allows for the description of XLM-based interfaces designed for the storage and transmission of the structured data to mobile client applications. The client applications used for the visualization of structured data together with DCM employ the ActionScript technology.

The data analysis and knowledge discovery system has a three-level architecture, which includes the following layers, in particular (1) the presentation layer, application server layer (2) and (3) the data layer (Fig.3).

**Presentation layer.** The personnel of power supply companies starts working with the system by opening the master portal through the HTTPS enquiry to the application server JBoss from the standard browser. After the authorization the user can chose the appropriate application for the operation. Afterwards the server transmits the page with JavaScript patches and Adobe Flex client applications. A flex client continues the operation with the server via the AMF protocol by means of the HTTPS Protocol.

**Application Server Layer.** The application server receives inquiries from the clients, performs the appropriate data-related operations and transmits the response data to the client. The application server acts both as the Web-server and intelligent data analysis server. It interacts with other servers via the API interface based on the Enterprise Java Beans (EJB) technology. The server performs computations using algorithms that require high processing power (aggregation, forecasting, and scenario analysis). To speed up computations when working with large data arrays the CUDA (Compute Unified Device Architecture) technology was realized, which allows for the data processing using a graphic video processor.

**Data Layer.** The data layer was realized using the Oracle DBMS server. The distributed tables of the Oracle DB keep both achieve and on-line data obtained from OPC servers of industrial controllers and other automation devices. Instrumental Data Feeds work in the multidimensional data layer and solve problems related to the intellectual data analysis and knowledge discovery [16]. This layer also uses the non relational (NoSQL) data storage system (Cassandra), which acts as the level-two cache for the ORM Hibernate and interacts with the application servers and Agent-Feeds.

Three main modules of data processing work at these layers:

1. Module SDF (Sensor Data Flow). It is used for the integration of new sensory data streams from connected objects of monitoring in multidimensional database.

2. Module EEMDPMD (Energy Efficiency Manager Dashboard). It is intended to prepare and generate reports and analytical graphs about changes in the parameters of urban energy efficiency.
3. Module ECLBPLB (Energy Consumption & Losses Breakdown). It is the tools for intellectual analysis of the sensor data include such components as:

   (a) Module of extraction of analyzed data slices from the hypercube;
   (b) Multidimensional data visualizer;
   (c) Aggregation mode editor.

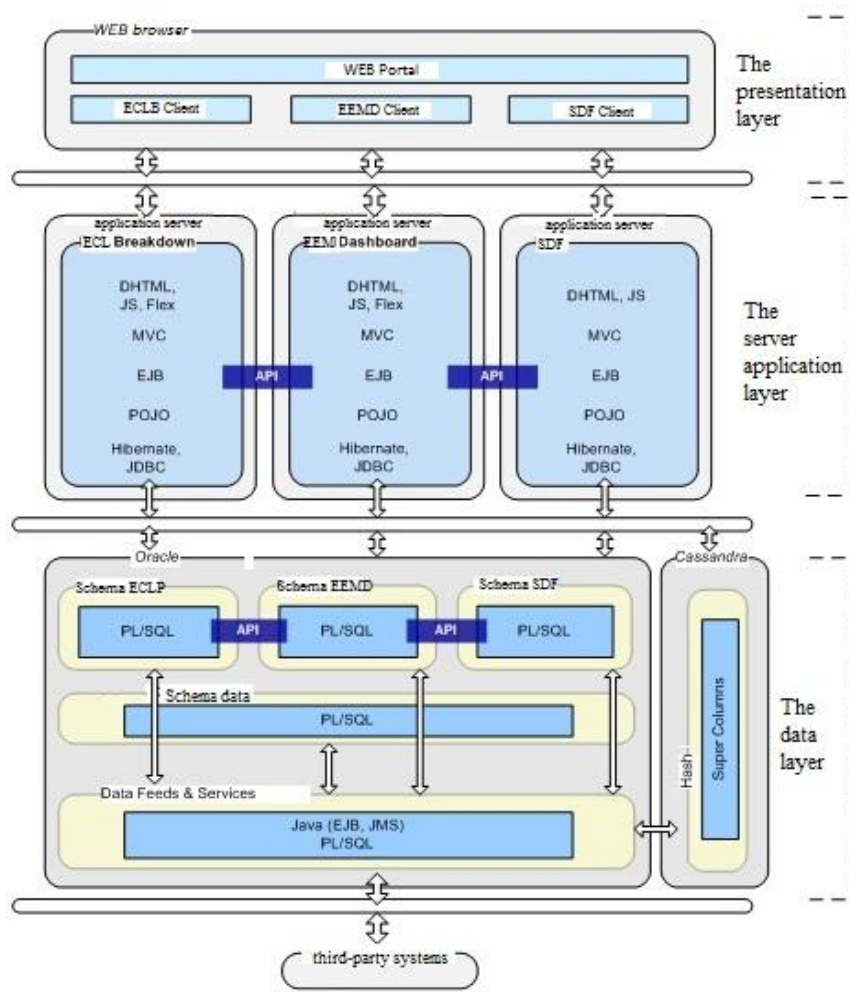

**Fig. 3.** - Fragment of the intellectual analysis system architecture

The multidimensional data visualizer allows for the presentation of the data for the analysis and the analysis data in the form of the structured hyper table. The hyper table combines the functionality of the classic table with a tree-like structure and the elements of structured data (knowledge) control and presentation on time scales. Actually, the hyper table presents an approach to the visualization of the results of intellectual analysis. It allows for the observation of a change in archive, current and forecast values of power efficiency indexes in the form of structured knowledge in time by shifting a courser on the time scale. To update data kept in the storage the data values contained in the hyper table are changed in real time for each selected interval, which is prescribed in compliance with the selected forecast horizon or the time horizon for the scenario analysis.

The aggregation mode editor is required to support the realization of the mechanism of multilevel aggregation during the selection of data for their presentation in the hyper table. The aggregation mode editor prescribes the type and content of the hyper table of archive and current indices and forecast energy indexes and defines visible value columns, amount and character of data pooling levels, color notations, etc. The editor includes its own set of instruments required for the operation of the set of aggregation parameters.

The data sources are placed on geographically spaced engineering communication facilities of the municipal power supply system, therefore the decision-making support and monitoring system should provide the users with geospatial analysis tools. The tools are realized by special analysis subsystem, which is realized on the ArcGIS 9 platform that operates with DCM downloaded from the standard Google Map и Yandex Map geo services, coordinate data from the relational database with the description of the objects of monitoring, data slices from the multidimensional storage and the intellectual analysis data from the hyper table. This instrument gives the users additional opportunities and ease of the visualization of spatially distributed information on the objects of monitoring and provides an opportunity to display on DCM the results of intellectual and spatial analysis in the form of plots, diagrams, tables and color differentiated cartographic zones.

To process large data arrays big companies use computational clusters consisting of the thousands of server sites and programs solutions to distribute problems between the sites on the basis of the model used for the programming of distributed computations (Map-Reduce). It is not economically appropriate to use such cluster structures for the municipal energy supply companies. Therefore the unified architecture of the computing device (CUDA) was implemented for the analytical processing of multidimensional storage data and the hyper table. According to the CUDA technology the processing of large data arrays is performed in real time in many graphic videocard processors Nvidia Quadro FX 5800 4GB (240 processors) based on the model of distributed computations (Map-Reduce) for cluster-type computer systems. The principle of similar computations is based on map functions and contraction (reduce) functions used for the functional programming.

At the first level of the main server runs distribution of input data between server nodes in the cluster network. At the second level server node data distributes between cores GPU. The role of the master server performs CPU node. Results of the function of each node processor accepted, aggregated and transmitted to a higher level from the operating unit to the master node of the cluster, where they are included in the full list. Set of libraries solves the problem of load distribution on the compute nodes of the server cluster. Toolkit that implements the Mapping function, preprocesses the input sensor data and generates a multiple pairs of "key-value", which, after groupings transferred to the toolkit Reduce function, which does work on groups of data pairs, retrieving data from them.

Let's consider the method of solving the problem of sensory data in the urban heating system. Suppose we have obtained from the database of archival data slice with thermal accounting devices for the last year. Need to find out which objects urban heating networks have a maximum energy consumption. In the first step the input list is received the master node of the cluster and distributed among the remaining nodes. In the second step, each node performs a predetermined Reduce function display on his part of the list, generating a pair, whose key is the name of the object, and the value - a value of energy consumption. Mapping operation work independently of each other and can be performed in parallel by all nodes in the cluster. The next step includes the master node on the resulting key key-value pairs and distributes the group with the same keys between nodes to perform the Reduce operation.

In reduce step all nodes in parallel perform a given function, which adds all the values for the input list, thus creating a single pair with the name of the monitoring object as the key and the number of occurrences of names in the original list as the value. After that, the master node receives data from the operating units and generates a result list, in which the records with the highest value and are the desired objects.

## 6      Conclusion

According to the system-synergetic approach the decision-making support and monitoring system functioning is based on the considered formalized strategies of the management of municipal power supply services on the basis of data collected from the engineering communication facilities. During the monitoring system design the level of the formalization of the management problem is defined by the availability of information on the technological processes of generation, transportation, consumption and utilization of energy carriers and also on the configuration of engineering networks and characteristics of individual objects of the power supply system. The development of the unified multidimensional storage system and system software will allow for the quality improvement, reduction of performance time and the cost of the realization of managerial decisions related to the satisfaction of needs of the population in energy resources.

At the present time some components of the monitoring system are realized by the heat supply service in the city of Kuznetsk, Penza region (Russia), in particular the municipal company "Gorteploset" and by the energy supply service in the city of Kharkov (Ukraine), in particular the Scientific Production Association "KHARTEP" and "INTEP". In particular a push was given to the consideration of issues related to the management hierarchy organization in the power and heat engineering, the formation of the composition and content of management problems at individual levels, coordination and harmonization of decision-making processes at different levels, increased control efficiency, development and implementation of information and telecommunication technologies for the acquisition and processing of the large arrays of sensor-based data.

The use of the decision-making support system for the power supply on the basis of the integration of the SCADA systems of different services into the unified system, the data integration in the multidimensional "cloudy" storage, the introduction of KDD and Data Mining for the monitoring of technological processes of energy generation, transportation, consumption and utilization guarantees the reduction of energy losses and the achievement of energy effectiveness.